# Toward a self-learned Smart Contracts

Ahmed Almasoud
*Faculty of Software*
*University of Technology, Sydney*
*Ultimo NSW 2007*
Ahmed.Almasoud@student.uts.edu.au

Maged M. Eljazzar, IEEE Member
*Computer Engineering Dept.*
*Faculty of Engineering, Cairo University*
*Giza 12613, Egypt*
mmjazzar@ieee.org

Farookh Hussain, IEEE Member
*Faculty of Software*
*University of Technology, Sydney*
*Ultimo NSW 2007*
Farookh.hussain@uts.edu.au

*Abstract* - In recent years, Blockchain technology has been highly valued and disruptive. Several research has presented a merge between blockchain and current application i.e. medical, supply chain, and e-commerce. Although Blockchain architecture does not have a standard yet, IBM, MS, AWS offer BaaS (Blockchain as a Service). In addition to the current public chains i.e. Ethereum, NEO, and Cardeno. There are some differences between several public ledgers in terms of development and architecture. This paper introduces the main factors that affect integration of Artificial Intelligence with Blockchain. As well as, how it could be integrated for forecasting and automating; building self-regulated chain.

*Index Terms*—Blockchain, *Smart contracts, Artificial intelligence, Distributed autonomous organization*.

## I. INTRODUCTION

Blockchain is the first truly stable decentralized global system. Bitcoin [1] is similar to other cryptocurrency in its concept, yet it is the most known and used case. The main aim of Bitcoin is to simplify the exchange of assets and services by offering a commonly trusted accepted good. Since it removes any intermediary approving authority (e.g. Visa or Mastercard) and offers trust between parities. The Blockchain transactions are not limited to that application. Thanks to smart contracts, they can represent digital assets between multiple parties. A smart contract is a software component encompass business logics and transactions that run on a Blockchain. Blockchain combined with smart contracts establish a new generation of decentralized system, and transactional applications based on trust between parities. Moreover, the accountability and transparency at their essence, streamlines the business processes and the handling legal constraints

With the rise of developing Ethereum smart contracts, a Blockchain became a decentralized general-purpose computer. Since it provides an open, decentralized, and fault-tolerant transaction technique. As a result, it attracted the attention of many financial corporations and public sectors. Which in turn sparked hundreds of chains regarding digital assets, medical field, Internet of Things (IoT), smart cities, supply chain, reputation systems and security services, money transferee and online payment [2-4].

Currently, several platforms offer the support of smart contracts. There are definitely some differences in architecture, consensus, programming language support, and privacy. But each one of them has its own supporting community and decentralized apps (DAPPS).

The main concept circulating around all of them refers to Blockchain as a database that stores all transactions grouped in blocks. In most cases, if a new transaction is created, the sender broadcasts it in the peer-to-peer (P2P) network to all the other nodes in the network. The transaction would then show as pending. As the nodes receive the transaction, they validate it and keep it in their transactional pools based on the consensus algorithm in the network. The Proof of Work protocol (PoW) is used in Bitcoin, Ethereum, and other chains. The state of the art behind PoW is to validate transactions in means of running predefined checks about the structure, and the actions in the single transaction. Miners (validation nodes) create a new block and include all, or some of the available transactions from their transactions pool. Then, the block is mined.

Even though the Blockchain technology has a great potential for the construction of future Internet systems, it is facing numerous technical challenges i.e. scalability, development, latency, and transactions per second. This paper introduces a review for smart contracts platforms and how AI could be integrated with smart contracts for rapid development. As Artificial Intelligence continues to make an extreme interaction with almost every faction of the society. Definitely, forecasting, optimization, and automating are the main AI contribution in several domains [5]. Which in turn reflects to self-learning smart contracts from the previous experience related to the chain. This paper is organized as follows: Section II provides a literature review of other smart contracts platforms. Section III compares differences between Blockchain features. Section IV includes AI factors in the Blockchain architecture. Integrating AI with smart contracts are presented and discussed in Section V. Finally, Section VI concludes the paper.

## II. LITERATURE REVIEW

Several platforms offer the support of smart contracts. Smart contracts are one of the most important aspects of Blockchain technologies, as well as the reason why Blockchains can be called a disruptive technology. It refers to the logic of the chain, and the increase in efficiency of our social structure over the time. Ethereum [6] is a first Blockchain-based ecosystem that provides a compatible environment to run smart contracts. It is a main standard in the world of smart contracts and has the major support community among other platforms. As a result, one of the Ethereum features is the Truffle development framework. It is a popular development framework for Ethereum that supports built-in smart contract compilation, linking, deployment, and unit testing.

NEO [7], also known as Chinese Ethereum, has some advantages over Ethereum since it is more scalable. It can process multiple transactions promptly, significantly minimizing wait times for its users. Another main advantage being, Ethereum is more bounded than NEO. Unlike the Solidity language in Ethereum, NEO smart contract can be coded directly by using most of high-level programming language. Besides, NEO developers' community provide plug-ins for these languages, which are used to compile high-level languages into instruction sets supported by NEO virtual machines. For now, the NEO community is growing. But it is still distant from Ethereum community, and its supporting frameworks due to its short age when comparing to Ethereum.

On the other side, LISK [8] avoids the standard barriers of entry in Blockchain technology. It is end-to-end Blockchain JavaScript project. LISK is based on the sidechain concepts [9]. Developers need to focus on developing D-Apps. Then, link D-Apps with main LISK ledgers. The main philosophy in LISK is to allow the developers to focus on the development with a common language, and leave the rest to the LISK chain. Through this method, developers can make use of Blockchain features without the headache of node infrastructure and configuration. This approach begins to spark and manipulate the Blockchain architecture in several projects. In doing so it decreases the administration headache for developers. Meanwhile in Cardeno [10], they are focused on building a group of design patterns, and engineering best practices. It builds a control layer to govern the usage of Smart contracts. The main aim of the layer is to recognize identity, and maintains a degree of separation from the currency it supports. Furthermore, in adding sidechains, Cardeno seeks for communication between several Blockchain and Sidechains. It is very simple for an ecosystem of Bitcoin to flow into Ethereum without needing to go through centralized exchanges.

At last, there is a Hyperledger project by Linux Foundation [11]. There are several projects introduced by Hyperledger community. Fabric is a permissioned Blockchain architecture taking the leading in Blockchain application. One of the differences in the structure between permissioned Blockchain and permissionless Blockchain is mining. In a permissioned Blockchain, nodes do the mining process. And in certain applications, the mining process may actually be delegated to authorized network members. The same concept would be applied to Verification of transactions and blocks.

## III. AI Factors in Blockchain Architecture

After reviewing several Blockchain architectures, it is important to understand why Blockchain architecture is significant. In addition, the main feature we are targeting from Blockchain may cause some latency, and scalability issues. Simply, Blockchain technology facilitates peer-to-peer transactions without any intermediary. One can consider Blockchain as simply a database that is public, distributed, is infinitely updated, and is secured by the cryptography. If the business requires these features; then a question can be raised. If listed participants are limited, should one build a public Blockchain or private Blockchain to serve to specific use-cases? This leads to huge concerns for selecting the best architecture that corresponds these use-cases.

Therefore, the major question is whether or not we'd need all Bitcoin Blockchain features, or could a distributed ledger be designed differently to fit an individual use case. In particular, confidentiality concerns are imposing the emergence of alternative Blockchain architectures referred to as 'private Blockchain'. Some distributed ledgers architecture sacrifices, in some decentralization to address some Blockchain shortage issues. For example, attaching Sidechains to Blockchain may handle some scalability issues. As a Blockchain architect, one can build Blockchains that require permission in order to read the information within the Blockchain. Thus, Limiting the network users who can transact on the Blockchain. And those who have the access to validate the data or read the data on chain or not. For an empowered smart contract with AI, several factors should be considered:

1- Mining.

Smart contracts are the mastermind of the chain. It gets executed by sending transactions between different parities. This transaction may include metadata to be used by the smart contract. In order to execute code in an Ethereum smart contract, each transaction requires a specific amount of gas. Which refers to the transaction cost. It is sent alongside the transaction along with the price in ether for each unit of gas. Gas is important for several security reasons. It prevents denial of service attacks (DoS), which may occur from the untrusted resources for executing purposeless contracts to make the network unresponsive. Besides, preventing infinite loops or the halting problem.

While in permissioned Blockchain, this may or may not involve 'proof of work' from the nodes. There are some politics around this, as there are those who consider private Blockchains that do not apply any of (PoW) (that is, Blockchains with no mining) to not be Blockchains at all, but simply shared ledgers.

2- Data on the Chain.

It is well known in public chain that each piece of information that one stores is duplicated into hundreds or more nodes (i.e. more than 100,000 in case of Bitcoin). Blockchain are not a good solution for storing data for substantial sizes, and medical health records as an example. But still, Blockchain is able ensure the transparency of transactions. One of the workarounds is to encrypt the data before adding it to the chain. Then, decrypt it in the application layer when the user needs to read it. For huge data, it is commonly stored in off-chain storage (i.e. IPFS, AWS) or by adding a fingerprint instead of the medical record. It then adds a hash address for data location.

By using this method, the more network users that take part in verification with respect to the current consensus algorithm, the more secured and decentralized the chain becomes!

On the other hand, the permissioned chain acts as traditional database, with some features of the Blockchain. Such as, permissions, multiple input validation, multiple copies, append-only writes and data transparency. The architect defines who has the capability to read and modify the data. For the health records, only few parities have an access to read the data based on rules and policies. That is why a Permissioned (Private) Ethereum or Stellar implementations would be a good option to manipulate the data. These kinds of Blockchain are similar to the public chain, but instead acts as a permissioned chain.

3- Network Actors.

The terms public, permissioned, and private refer to the network user. In other words, the person who would be able to operate a node, access to the data and validate the transaction. Public Blockchain refers to decentralized system, where anyone can be a network user. In some cases, there is a little restriction for low resourced users. On the other hand, private Blockchains are operated by known entities such as employees in a company. They have the value of immutability and efficiency over alias and transparency. Mainly laying in the area of a specific industry. In supply chain, the Blockchain enables goods to be tracked from raw material to completed form during the supply chain cycle.

4- Blockchain Integrity

Current chain has some integration with several concepts (i.e. Sharding, state channel, Master node architecture, etc.). Sharding is a common term in the database. It is a type of database partitioning to divide large database into smaller and faster parts known as data shards. While the target of using State Channels is helpful in retaining privacy, as it allows the recording and execution of transaction between two parties on an off-chain platform, the reference for such transactions is saved in relevant Blockchain system after their successful completion. However, such transactions are useful through State Channels only for smaller payments involving frequent sending, and receipt of payments between two parties. The system does not support in case large amount transactions are involved.

IV. INTEGRATING SMART CONTRACTS WITH AI

The aim of smart contracts is the automation of business by reducing the usage of paperwork and by eliminating intermediaries (e.g., lawyers). Thus, decreasing costs. Wills and all sorts of Legal documents can be processed using smart contracts such as NEO or straits. AI can be integrated with smart contracts through rules and policies assigned to the chain. Leading to the integration of Self-learning artificial intelligence and smart contracts. Assuring that they are legally binding, and allowing their adaption in case of environmental changes.

The deep understanding of rules and policies in the chain, and the analysis of where AI should come in enables the creation and execution of such complex smart contracts. Therefore, making them more effective and learnable. The more data AI has, the better it can predict outcomes. For example, in contract negotiation in supply chain, AI could act as a recommendation system. It can analyze archived contracts to view how parties negotiated in the past. The AI recommendation engine would then be able to suggest the types of language and clauses that are most likely to secure agreement between different parities in the chain. Afterwards, AI can analyze past contracts to identify factors that are not previously considered, and incorporate them into future contracts. This section explains how AI could be integrated with smart contracts in terms of (Network users, data on the chain, and mining).

1- Smart contracts cost.

It is well known that AI perquisites computation resources. In Ethereum, deploying a contract has a minimum fee of 32000 gas, and 200 gas per byte of the source code. In addition to calling and storing data cost. The fee is 20k gas to store a 256-bit word. A kilobyte is thus 640k gas. The basic idea behind storing data is gas prices based on Oracle's and miners. It will go up exponentially as ETH appreciates. But when it comes to AI one should consider storing and manipulating a lot of data. However, this may be very costly. In comparison with storing data off-chain, cost could be estimated by the following equation:

$$Gas\ Used * Gas\ Price * USD/ETH$$

For example, 32000 Gas with average gas price (8 Gwei) cost $0.18074. This number seems very small, but in repeating these smart contracts in every transaction would be costive.

2- Rules and policies.

It is important to elaborate the rules that would be applied through AI in the chain. In some cases, it acts as a recommendation engine (i.e. a recommendation for tracing items in supply chain). While in other cases, it acts as comprehensive data analytic for all transactions in the chain. Both of these scenarios have a different approach. a) Real time Recommendation engine

To have a recommendation engine, a historical data is analyzed. Rules and policies are defined based on the data to match business logic. Working on off-chain data and caching the previous experience of smart contracts in an off-chain storage, assists in training a model with several scenarios in real time. Training a model to be able to act as a recommendation engine in real time. This approach allows developers to only need use of read data from smart contracts with no cost in most public Blockchain. Accordingly, linking these API (Application Program Interface) with data stored off-chain. Finally, writing smart contracts consuming these API in real time.

Some of public Blockchain offers supporting side-chains transactions, Sharding assists in executing small transactions in shorter time.

b) Analytic Dashboard

According to the network users, all data stored in the Blockchain are analytics based on a set of rules and policies, in supply chain use-case. The manager needs to inspect all production lines, point of sales, and workers' activity. Besides, defining a set of variables to automate several tasks. These variables are translated into predefined rules, which in turn are reflected into smart contracts to automate planning as shown in figure 1.
The state of the art behind smart contracts is setting conditions.

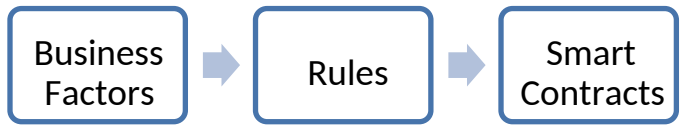

Figure 1. Show the process of self-learning smart contracts.

c) Self-learned Smart Contracts

Smart contract implies that event A is deemed incomplete until event B has taken place. To make smart contracts smarter, one can follow several approaches. Firstly, defining several rules to meet several application scenarios. These rules may not be active in the beginning with working on the chain, but it will work in time. To make rules smarter, one makes use of the important factors on the chain and sets rules based on the data. In such cases, when increasing the number of network users in the chain Event C will take place. When there are huge numbers of users, more transactions are related to Event C than D. A specific action would occur (i.e. new smart contracts would define to resolve a shortage in the network or add a new feature in the logic of the chain). Lastly, from the experience of the chain, new rules could be defined. This could be acting as analyzing the network participants' behavior towards transactions, defining a new set of rules based on these behaviors to achieve a specific target.

The last approach commonly known as DAO (decentralized autonomous organization) refers to an organization that exists as autonomous code on the Blockchain with no physical presence or employing people. The code is embedded into a smart contract that is executed once its conditions are fulfilled. Most of these conditions attempt to exclude the reliance of the transaction participants from any kind of centralized system or the human factor influence. Yet, it guarantees the automation, transparency, and safety of the entire process.

When coding DAO, it is a fully autonomous system. Hence, the limitation of functionality of smart contracts decides the degree of automation. In such case, creating new digital assets, and adding more smart contracts to handle specific complain in the network based on participants' transactions; as shown in figure 2. Many transactions are automated on the chain through smart contracts. In such a case, one needs to define new digital assets that depend on previous transactions. Defining a set of rules to edit shortage in a previous smart contract (i.e. updating the version of smart contracts). For example, due to new policies, event C will replace event B if event A took place.

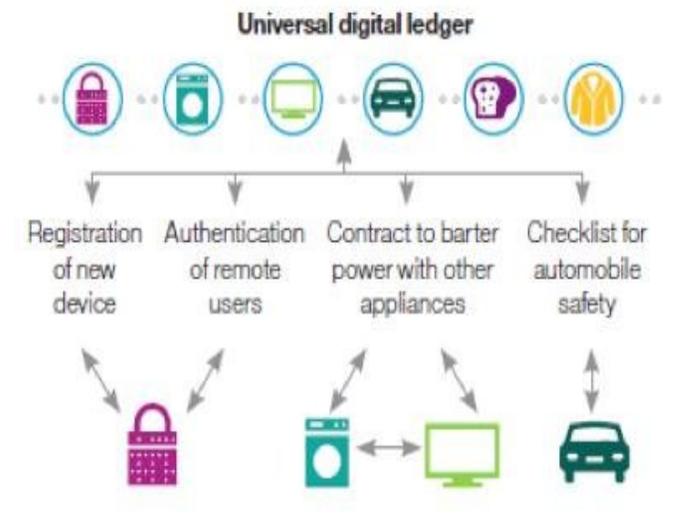

Figure 2. Show a smart city transaction in a public chain.

V. CONCLUSION AND FUTURE WORK

Artificial intelligence and Blockchain are disruptive technologies. Blockchain is conquering several fields as a new decentralized system offering some features (i.e. encryption, trust, and data transparency). On the other side, AI became a crucial tech player in most of technologies. This paper introduces the main factors that impact the integration of AI with Blockchain. Until now, Blockchain does not have any standards; it is important to keep eye on the smart contracts cost, transactions fees, and storage on chain or off chain.

To obtain the best practices, storing data off chain would be a better option for real time recommendation engine to prevent latency. Further, it would be more cost effective. On the other side, smart contracts are mainly based on a set of conditions, defining rules and policies that meet these conditions. More business automation could lead to some automation. While learning from experience transactions types is a necessity to define new smart contracts with new business scenarios based on the chain network.

In this work, fewer comparisons have been introduced between several Blockchain architectures and the outcome of these factors on the integration with AI. In future work, we will apply a use case scenario and test the performance on several platforms (NEO, Ethereum, Straits). Comparison should include the latency, ease of development, and any restriction in the network.